\documentclass{aa}  

\usepackage{graphicx}
\usepackage{txfonts}
\begin{document}

   \title{Submillimeter H$_2$O masers in water-fountain nebulae\thanks{This publication is based on data acquired with the Atacama Pathfinder Experiment (APEX). APEX is a collaboration between the Max-Planck-Institut fur Radioastronomie, the European Southern Observatory, and the Onsala Space Observatory}}


   \author{D. Tafoya\inst{1,2}
          \and R. Franco-Hern\'andez\inst{3,4} 
          \and W. H. T. Vlemmings\inst{1} 
           \and A. F. P\'erez-S\'anchez\inst{5} 
           \and G. Garay\inst{4} 
          }

   \institute{Chalmers University of Technology, Onsala Space Observatory SE-439 92 Onsala, Sweden\\
              \email{d.tafoya@crya.unam.mx}
         \and
             Centro de Radioastronom\'\i a y Astrof\'\i sica, UNAM, Apdo. Postal 3-72 (Xangari), 58089 Morelia, Michoac\'an, M\'exico\\
         \and 
              Instituto de Astronom\'\i a y Meteorolog\'\i a, Universidad de Guadalajara, Avenida Vallarta No. 2602, Col. Arcos Vallarta, C.P. 44130 Guadalajara, Jalisco, M\'exico
         \and
            Departamento de Astronom\'\i a, Universidad de Chile, Casilla 36-D, Santiago, Chile \\
         \and
             Argelander Institute for Astronomy, University of Bonn, Auf dem H\"ugel 71, D-53121 Bonn, Germany 
             }

   \date{Received; accepted}

 
  \abstract
  {We report the first detection of submillimeter water maser emission toward water-fountain nebulae, which are post-AGB stars that exhibit high-velocity 
  water masers. Using APEX we found emission in the ortho-H$_{2}$O~(10$_{29}$$\rightarrow$9$_{36}$) transition at 321.226~GHz toward three sources: IRAS~15445$-$5449, 
   IRAS~18043$-$2116 and IRAS~18286$-$0959. Similarly to the 22~GHz masers, the submillimeter water masers are expanding with a velocity larger than that of the OH masers, 
  suggesting that these masers also originate in fast bipolar outflows. In IRAS~18043$-$2116 and IRAS~18286$-$0959, which figure among the sources with the fastest water 
  masers, the velocity range of the 321~GHz masers coincides with that of the 22~GHz masers, indicating that they likely coexist. Towards IRAS~15445$-$5449 
  the submillimeter masers appear in a different velocity range, indicating that they are tracing different regions. The intensity of the submillimeter masers is comparable to 
  that of the 22~GHz masers, implying that the kinetic temperature of the region where the masers originate should be T$_{\rm k}>1000$~K. We propose that the passage of two 
  shocks through the same gas can create the conditions necessary to explain the presence of strong high-velocity 321~GHz masers coexisting with the 22~GHz masers in the same region.}

   \keywords{Masers -- Stars: AGB and post-AGB -- Submillimeter: stars }

   \maketitle
%

\section{Introduction}

Water-fountain nebulae are thought to represent a group of post-asymptotic giant branch (post-AGB) stars where the formation of the bipolar and multipolar 
morphologies observed at the planetary nebula phase has recently been triggered. In these sources the water masers trace high-velocity collimated outflows. 
It has been proposed that they appear as a result of an episode of collimated mass-loss in the post-AGB phase that modifies the morphology of the circumstellar 
envelope (CSE; Likkel et al. 1988).  Up to date, the water maser emission that has been observed in these objects arises from the 
$J_{K_{a}K_{c}}=6_{16}$$\rightarrow$5$_{23}$ transition at $\sim 22$~GHz. Since the physical conditions to invert this transition require high densities 
($\sim 10^{8} - 10^{10}$~cm$^{-3}$) and kinetic temperatures T$_{\rm k} \sim 200 - 2000$~K (Yates et al. 1997), it has been proposed that this emission arises 
from shocked dense material that has been swept up by the collimated wind that pierces into the CSE (e.g. Imai 2007). In order to better constrain the physical 
conditions of the gas within which the masers originate and to reveal the characteristics of the collimated wind that provides the energy, observations of other 
transitions of water maser emission toward these type of sources are necessary.

In addition to the H$_{2}$O~(6$_{16}$$\rightarrow$5$_{23}$) line, several water maser lines, most of them at submillimeter wavelengths, have been 
detected toward star-forming regions and late-type stars (Menten et al. 1990a, Menten et al. 1990b, Melnick et al. 1993, Patel et al. 2007, 
Menten et al. 2008). Some transitions of the submillimeter water masers have upper levels with energies above the ground state higher than that of the 
22~GHz masers, for which $E/k=643$~K. Particularly, the upper level of the 321~GHz water maser transition has an energy $E/k=1861$~K above the 
ground state, consequently, these masers trace dense gas with relatively high temperatures. In the star-forming region Cepheus A, Patel et al. (2007) found 
321~GHz water masers that are aligned with the bipolar outflow seen in this source. This indicates that these submillimeter masers are arising in 
material that has been shocked and heated up by the jet. In late-type stars the physical conditions to efficiently pump the 321~GHz water masers
(n$_{\rm H_{2}}=8\times10^{8}-4\times10^{9}$~cm$^{-3}$; T$_{\rm k}>1000$~K; Yates et al. 1997) are found in the inner CSE,  with strongest 321 GHz emission 
confined to a distance closer to the star than the region traced by strong 22 GHz masers (Humphreys et al. 2001). The 22~GHz water masers 
in water-fountain nebulae are associated with high-velocity collimated outflows. 

We searched for submillimeter H$_{2}$O maser emission toward a group of water-fountain nebulae that exhibit relatively strong 22~GHz water maser emission. 
In this letter we report the first detection of  H$_{2}$O~(10$_{29}$$\rightarrow$9$_{36}$) maser emission at 321.226~GHz toward three of these sources. 
\section{Observations}

The observations were carried out in two epochs using the 12m Atacama Pathfinder Experiment (APEX) telescope. We made single pointing observations
 using the heterodyne receiver APEX-2 centred at the rest frequency of the 
H$_{2}$O~(10$_{29}$$\rightarrow$9$_{36}$) transition, $\nu_{0}=321.22568$~GHz. The system temperature ranged from 260~K to 430~K, 
except when we observed the source IRAS~18450$-$0148 for which T$_{\rm sys}\sim1100$~K. The calibration was performed using the chopper wheel 
technique. The conversion factor from antenna temperature to flux density at $\sim 345$~GHz is 41~Jy~K$^{-1}$. We reduced the data using the package 
CLASS of GILDAS, averaging individual spectrum of each scan and subsequently subtracting a fitted base-line of first order to the emission-free channels. We 
smoothed the resulting averaged spectra to a final velocity resolution of $\sim 0.6$~km~s$^{-1}$. On May 4th 2013 we also carried out observations 
of the para-H$_{2}$O~(5$_{15}$$\rightarrow$4$_{22}$) transition, $\nu_{0}=325.1529$~GHz, with APEX. However, we did not detect emission in any of 
the sources above an rms noise of 2~Jy.

\section{Results and discussion}

We observed seven water-fountain nebulae with relatively strong 22~GHz maser to search for submillimeter H$_{2}$O maser emission. 
The three sources in which we detected 321~GHz water emission are listed in Table~1. Their spectra are shown in Figure~1. Although 
the emission showed variability by a factor of up to $\sim$10 in 
some spectral features, it was detected toward the three sources in both observation epochs. The sources with no detection 
are listed in Table~2.  The 321~GHz water emission from IRAS~18043$-$2116 and IRAS~18286$-$0959 consists of clusters of spectral lines spread over 
$\gtrsim 200$~km~s$^{-1}$. The narrow width of the lines suggests that the emission is indeed amplified by the maser effect. The
emission from IRAS~15445$-$5449 shows a broader line width. However, thermal emission of the H$_{2}$O~(10$_{29}$$\rightarrow$9$_{36}$) 
transition has not been reported before. Additionally, the spectrum also exhibits narrow spectral features typical of maser emission.

In the following subsections we describe the characteristics of the maser emission from each source.  
 \begin{table*}
\caption{\label{Tab1}Sources with detected 321~GHz water maser emission}
\begin{center}
\begin{tabular}{lcccccc}
\hline\hline
Source&RA(J2000)&Dec(J2000)&line peak&rms&line peak&rms\\
&&& (epoch 1)&(epoch 1)&(epoch 2)&(epoch 2)\\
\hline
IRAS name&h~~~m~~~s&~~$\circ$~~~$\prime$~~~$\prime\prime$ & Jy&Jy&Jy&Jy\\
\hline
15445$-$5449&15 48 19.37&$-$54 58 21.2&1.6\tablefootmark{a}&0.4\tablefootmark{a}&1.6\tablefootmark{a}&0.3\tablefootmark{a}\\
18043$-$2116&18 07 21.10&$-$21 16 14.2&8.5&0.9&4.2&0.7\\
18286$-$0959&18 31 22.93&$-$09 57 19.8&25.2&1.0&11.1&0.7\\
\hline
\end{tabular}
\tablefoot{Line peak and rms noise values for a spectral resolution of 0.6 km~s$^{-1}$.\\
\tablefoottext{a}{Values for a spectral resolution of 2.6 km~s$^{-1}$.}
}
\end{center}
\end{table*}

\subsection{IRAS~15445$-$5449}

\begin{figure}
   \centering
   \includegraphics[width=\hsize]{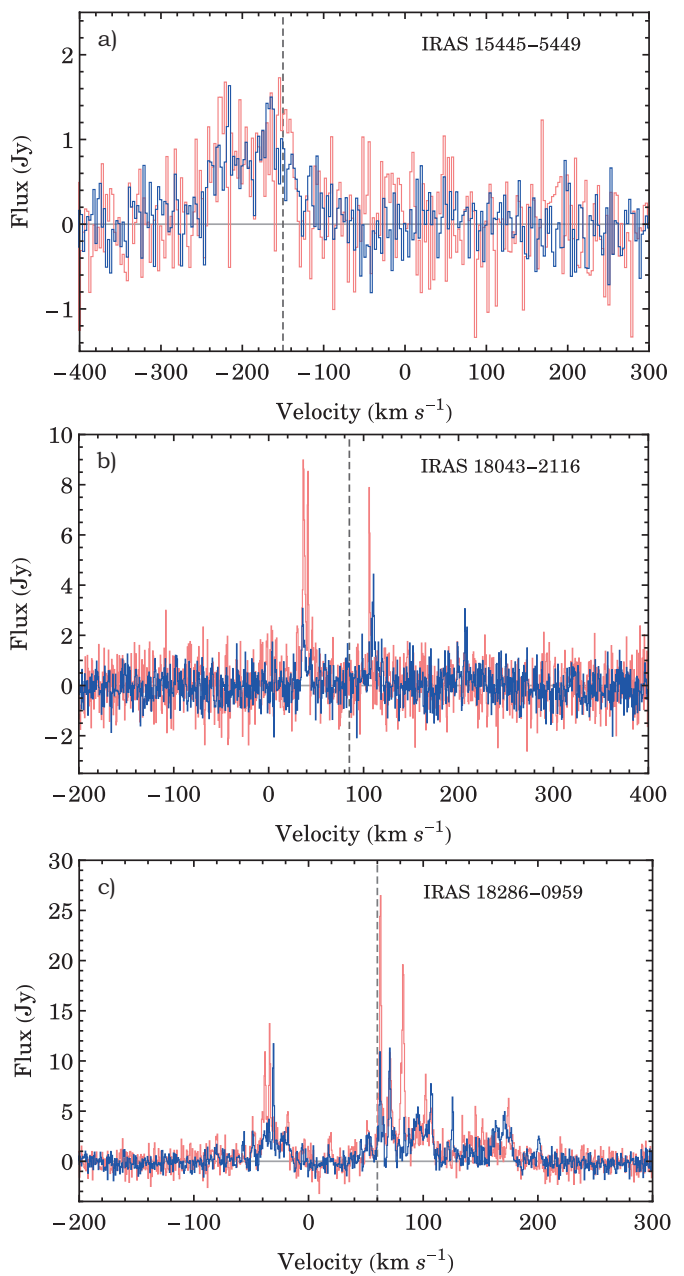}
      \caption{Spectra of the H$_{2}$O~(10$_{29}$$\rightarrow$9$_{36}$) maser emission in three water-fountain nebulae. 
      The red line indicates the spectrum obtained in the epoch May 15th 2013 and the blue line indicates the spectrum obtained in 
      the epoch July 6th 2013. The systemic velocity reported in the literature is indicated with a vertical dashed line. Note that 
      the velocity and flux ranges are different for each source.}
         \label{fig_spectra}
   \end{figure}
The 321~GHz water emission in this source comprises a broad spectral feature that covers the velocity range 
from $\sim-240$~km~s$^{-1}$ to $\sim-110$~km~s$^{-1}$ (see Figure 1a). It did not exhibit significant variability between 
the two epochs of our observations. In this source, the 22~GHz maser emission reported by Deacon et al. (2007) and P\'erez-S\'anchez et al. (2011) appears over the velocity 
range from $\sim -150$~km~s$^{-1}$ to $\sim -75$~km~s$^{-1}$. In addition, the OH maser emission consists of a broad spectral feature that 
covers the velocity range from $\sim -200$~km~s$^{-1}$ to $\sim -110$~km~s$^{-1}$ (Deacon et al. 2007). Thus, the 321~GHz H$_{2}$O maser 
emission has a velocity range in better agreement with that of the OH maser emission than with that of the 22~GHz water masers.  

Since the OH maser emission does not show the typical double peak characteristic of AGB stars, the systemic velocity of this source is not well known. 
Deacon et al. (2007) adopted a systemic velocity equal to the midpoint of the OH profiles, near a velocity of v$_{\rm LSR}=-$150~km~s$^{-1}$ (see Figure 1a). This 
would imply that most of the submillimeter masers are blue-shifted with respect to the source. Thus, they might be tracing the blue-shifted part of the bipolar outflow seen 
in this source (Legadec et al. 2011; P\'erez-S\'anchez et al. 2013). Another possibility is that, given the double-peak of the profile of the 321~GHz line emission, the systemic 
velocity is closer to v$_{\rm LSR}=-$185~km~s$^{-1}$. This would indicate that both the OH and the 22~GHz masers are red-shifted with respect to the star. 

\subsection{IRAS~18043$-$2116}

The velocity range covered by the 321~GHz water emission extends from $\sim 30$~km~s$^{-1}$ to $\sim 210$~km~s$^{-1}$, 
which falls within the velocity range of the 22~GHz masers reported by Walsh et al. (2009). In this source the 321~GHz maser emission exhibits at least three spectral 
features centred at around v$_{\rm LSR}\sim 40$~km~s$^{-1}$, $\sim 110$~km~s$^{-1}$ and $\sim 210$~km~s$^{-1}$, respectively. The last one only appears in the July 6 
observation epoch, indicating the high variability of this emission (see Figure 1b). Comparing our spectrum with those obtained by Deacon et al. (2007), Walsh et al. (2009) 
and P\'erez-S\'anchez et al. (2011) we note that, in general, the intensity of the submillimeter masers is similar to that of the 22~GHz masers. The systemic velocity 
of this source is around v$_{\rm LSR}=85$~km~s$^{-1}$ (Sevenster \& Chapman 2001), implying that the submillimeter masers arise in gas expanding with velocities of at least 
 125~km~s$^{-1}$.

\subsection{IRAS~18286$-$0959}
This source exhibits a very rich 22~GHz water maser spectrum with emission spread over a velocity range $\gtrsim 200$~km~s$^{-1}$ and distributed in 
a double-helix pattern (Deguchi et al. 2007; Yung et al. 2011). The 321~GHz water emission also 
extends over a wide velocity range from $\sim -50$~km~s$^{-1}$ to $\sim 200$~km~s$^{-1}$, similar to that of the 22~GHz masers. The 321~GHz spectrum 
shows a complex ensemble of spectral lines that seem to be grouped in at least three broad features (see Figure 1c). There is weak submillimeter water emission at 
$\sim 200$~km~s$^{-1}$ with no counterpart 
in the 22~GHz emission. The brightest 321~GHz maser emission correspond with the strongest lines at 22~GHz, 
although the spectra changed significantly between the two epochs of observation. The systemic velocity of this source has been adopted as v$_{\rm LSR}=50$~km~s$^{-1}$ 
by Yung et al. (2011) and v$_{\rm LSR}=60$~km~s$^{-1}$ by Imai et al. (2013). Thus, the submillimeter masers exhibit expansion velocities of at least $\sim140$~km~s$^{-1}$.
\begin{table}
\caption{\label{Tab1}Sources with non-detected 321~GHz water maser emission}
\begin{center}
\begin{tabular}{lccc}
\hline\hline
Source&RA(J2000)&Dec(J2000)&rms noise\\
\hline
IRAS name&h~~~m~~~s&~~$\circ$~~~$\prime$~~~$\prime\prime$ & Jy\\
\hline
15103$-$5754&15 14 18.45&$-$58 05 20.3&0.8\\
16342$-$3814&16 37 39.91&$-$38 20 17.3&0.9\\
18113$-$2503&18 14 27.27&$-$25 03 00.5&0.7\\
18450$-$0148&18 47 40.97&$-$01 44 55.4&2.6\tablefootmark{b}\\
\hline
\end{tabular}
\tablefoot{rms noise for a spectral resolution of 0.6 km~s$^{-1}$.\\
\tablefoottext{b}{Value for a spectral resolution of 2.5 km~s$^{-1}$.}
}
\end{center}
\end{table}

\subsection{The origin of the 321~GHz water masers}

Theoretical works indicate that the excitation mechanism of the astronomical water masers must be collisional (de Jong 1973, Deguchi 1977, 
Cooke \& Elitzur 1985; although see Yates et al. 1997). These works also show that in addition to the $J_{K_{a}K_{c}}=6_{16}$$\rightarrow$5$_{23}$ 
transition several other lines should 
be amplified by maser effect. Among the predicted maser lines was the H$_{2}$O~(10$_{29}$$\rightarrow$9$_{36}$) at $\sim321$~GHz, which was 
detected for the first time by Menten, Melnick \& Philips (1990) in a group of sources that also exhibit 22~GHz masers. Neufeld \& Melnick (1990) 
estimated the physical conditions under which these two masers can coexist in the same volume of gas. In general, they showed that the values of density and 
temperature under which the 321~GHz masers are pumped are more constrained than those for the 22~GHz masers. These results were confirmed 
and extended in a more comprehensive work by Yates et al. (1997), who found that strong inversion of the 321~GHz transition is optimized in kinetic 
temperatures, T$_{\rm k} > 1000$~K, and densities, ($n_{\rm H_{2}} = [4-6]\times 10^{8}$~cm$^{-3}$). At higher densities, the 321~GHz emission can 
be stronger than the 22~GHz maser.

The 321~GHz maser emission in the detected water-fountain nebulae shows spectral characteristics that indicate that it could have 
different origins. While in IRAS~15445$-$5449 the 321~GHz maser emission does not coincide with the 22~GHz masers, 
in IRAS~18043$-$2116 and IRAS~18286$-$0959 the 321~GHz masers cover velocity ranges similar to those of the 22~GHz masers. 
This suggests that the 321~GHz masers in IRAS~15445$-$5449 are originating in a region different from that where the 22~GHz 
are located, possibly at a distance closer to the star. However, we note that the velocity extent of the submillimeter masers, 
$\sim130$~km~s$^{-1}$, is too large to be originating in the fossil CSE. We speculate that the water molecules in this source could be excited 
along the vicinity of the non-thermal jet detected by P\'erez-S\'anchez et al. (2013). In the other two sources, given that both masers span similar 
velocity ranges and that several spectral features appear at the same velocity, it is likely that the 321~GHz and 22~GHz masers are originating 
in the same gas. 

The models that explain the water maser emission in evolved stars assume that the emission originates in the expanding circumstellar envelope
created by the massive wind at the end of the AGB phase (Cooke \& Elitzur 1985; Neufeld \& Melnick 1990). However, the origin of the H$_{2}$O maser 
emission in water-fountain nebulae is associated to fast collimated outflows that interact with the slowly expanding CSE. Therefore, it is more appropriate 
to interprete the water maser emission in a similar way to that of the star-forming regions. Elitzur, Hollenbach \& McKee (1989) explained the water maser 
emission in star forming regions as the result of the passage of a dissociative shock (v$_{\rm s}\geq 50$~km~s$^{-1}$) through the interstellar medium (see also 
Hollenbach, Elitzur \& McKee 2013 and references therein). Behind the shock, a layer of high-density gas with a temperature of $\sim 400$~K forms, where 
the conditions for 22~GHz maser emission are optimal in that model. Neufeld \& Melnick (1990) showed that under the physical conditions of the post-shock region described by 
Elitzur, Hollenbach \& McKee (1989) 321~GHz emission can be produced with a luminosity ratio $L_{p}$(22~GHz)/$L_{p}$(321~GHz)~$\gtrsim 5$. 
They also suggested that values of the ratio smaller than five could be attained with slower non-dissociative shocks that would heat the molecules to 
temperatures up to T$_{\rm k}=1000$~K (Kaufman \& Neufeld 1996).   

For the case of the water-fountain nebulae presented in this work, the expansion velocity of the 22~GHz and 321~GHz H$_{2}$O masers is $\gtrsim 100$~km~s$^{-1}$, 
thus the shock must be dissociative. According to the model proposed by Elitzur, Hollenbach \& McKee (1989), when the shocked material cools down, H$_{2}$ and H$_{2}$O 
molecules form in gas that is maintained at T$_{\rm k}=400$~K by the energy liberated when the H$_{2}$ molecules are released from the dust grains. As mentioned 
above, for this temperature the luminosity ratio $L_{p}$(22~GHz)/$L_{p}$(321~GHz) is expected to be $\gtrsim5$. Comparing the intensities of the 22~GHz masers 
(see Deacon et al. 2007; Walsh et al. 2009; P\'erez-S\'anchez et al. 2011; Yung et al. 2011) and the 321~GHz masers of IRAS~18043$-$2116 and IRAS~18286$-$0959 it 
can be seen that the luminosity ratio is close to unity. This implies a kinetic temperature T$_{\rm k}>1000$~K for 
the gas (Neufeld \& Melnick 1990; Yates et al. 1997).  But this temperature should be indicating the presence of a relatively slow non-dissociative shock, in contradiction with the 
relatively high velocity of the masers. Therefore, to explain the coexistence of strong 321~GHz masers with 22~GHz masers, there should be a mechanism that 
accelerates the gas to the observed high velocities (v$_{\rm exp}\gtrsim 100$~km~s$^{-1}$) and that favours the creation of water molecules while maintaining 
the gas at high temperatures (T$_{\rm k}>1000$~K). 

The presence of high-velocity strong 321~GHz masers coexisting with 22~GHz masers in the same region could be explained if we invoke the passage of two 
shocks with different speeds through the same material. The first shock would be due to the collision between a fast collimated wind and the slowly expanding 
CSE, which produces a J-type dissociative shock (see above). If the post-shock density is much higher than that of the pre-shock gas then the velocity of the shocked gas would
be very low in the frame of reference of the shock. Thus, in the frame of reference of the star the shocked gas, within which the 22~GHz water masers originate, would move 
almost at the same speed of the shock (v$_{\rm exp}\sim 100$~km~s$^{-1}$). Subsequently we consider  a collision between the fast collimated wind with the shocked 
material. Since the shocked material is already moving at high speeds, the collision occurs at a slower relative velocity. This produces a slower C-type non-dissociative 
shock, which raises the temperature of the shocked gas to a higher value, where the 321~GHz transition is inverted more efficiently (Kaufman \& Neufeld 1996). The 
321~GHz water masers would move at the same velocity as the 22~GHz masers, which is the velocity of the shocked gas. If the temperature of the shocked gas reaches 
values T$_{\rm k}>1500$~K, then the gain coefficient of the 22~GHz transition becomes smaller 
and the 321~GHz masers become dominant (Yates et al. 1997). If the C-type shock does not occur or if there is an efficient cooling mechanism in the shocked material 
then the 321~GHz maser emission will be negligible. Interestingly, the velocity range and strength of the 22~GHz maser emission of the sources with no detected submillimeter 
water masers are not fundamentally different from those in which submillimeter masers were detected. This may be due to the lack of the second slower shock, or to the 
fact that the gas has already cooled down in the sources with no submillimeter maser detected. In this regard, the sources with strong high-velocity 321~GHz 
water masers represent a subclass of post-AGB stars that could be referred to as {\it hot-water fountain nebulae}. It is clear that our observations pose a challenge for the 
current water maser excitation models. 

It should be pointed out that our assumption of the coexistence of 321~GHz and 22~GHz in at least two of the sources is based on the correspondence of 
the velocity ranges of the maser emission at both frequencies. In order to confirm this assumption, simultaneous interferometric images of the maser emission 
at both frequencies are required. Patel et al. (2007) obtained maps of the 321~GHz and 22~GHz masers in the star-forming region Cepheus A. They found that 
while the 22~GHz masers are mainly tracing an equatorial structure, the 321~GHz maser are aligned in the perpendicular direction, parallel to the bipolar jet. 
In the case of the water-fountain nebulae presented in this work, the high expansion velocity of the 321~GHz masers strongly suggest that they are tracing 
the same structures as the 22~GHz masers. Future observations with ALMA will reveal the spatial distribution of the 321~GHz masers in water-fountain nebula 
that will allow us to compare with that of 22~GHz masers, which will help us to elucidate their origin.

\section{Conclusions}

We detected for the first time submillimeter H$_{2}$O maser emission in three water-fountain nebulae. In a similar way to the 22~GHz water masers, the 
submillimeter water masers extend over a velocity range broader than that of the OH maser emission, indicating that they are also tracing fast bipolar outflows. 
In general the kinetic temperature of the gas where the maser emission is originating is required to be T$_{\rm k}>1000$~K. Such kinetic temperature can 
be due to a slow non-dissociative shock. However, the large velocities of the H$_{2}$O masers indicates the passage of a faster dissociative shock. 
To explain our observation we propose that the passage of a fast shock accelerates the material and the water molecules are created behind it. Subsequently, 
a second slower shock heats up the shocked material to the temperates necessary to invert the 321~GHz efficiently. Higher angular observations with ALMA 
and more theoretical models are required to fully understand the presence of high-velocity 321~GHz masers in the subclass of hot-water fountain nebulae.

\begin{acknowledgements}
The authors acknowledge that Yolanda G\'omez played a crucial part in the initial formulation of this project. The authors thank Karl Torstensson for his valuable 
help with the observations at the APEX site. The authors also thank to the anonymous referee for his/her constructive comments and suggestions, helping to 
improve the manuscript. W.V. acknowledges support from Marie Curie Career Integration Grant 321691. G.G. acknowledges support from CONICyT project PFB-06.
\end{acknowledgements}


\end{document}